\documentclass[journal,12pt,onecolumn,draftclsnofoot]{IEEEtran}

\usepackage{amsmath}
\usepackage{graphicx}
\usepackage{amssymb}
\usepackage{mathrsfs}
\usepackage{mathtools}
\usepackage{makecell,multirow}

\usepackage[ruled,vlined]{algorithm2e}

\usepackage{bm}
\usepackage{stfloats}

\usepackage{cite}

\usepackage{subfigure}
\usepackage{color}

\newcommand{\MyAlterDel}[1]{}

\usepackage{booktabs}
\usepackage{multirow}
\usepackage{array}

\usepackage{flushend}

\begin{document}
\bstctlcite{Ref-MEC-TWC:BSTcontrol}
\makeatletter
\def\bstctlcite{\@ifnextchar[{\@bstctlcite}{\@bstctl
cite[@auxout]}}
\def\bstctlcite[#1]#2{\@bsphack
\@for\@citeb:=#2\do{%
\edef\@citeb{\expandafter\@firstofone\@citeb}%
\if@filesw\immediate\write\csname #1\endcsname{\s
tring\citation{\@citeb}}\fi}%
\@esphack}
\makeatother

\title{Data Age Aware Scheduling for Wireless Powered Mobile-Edge Computing in Industrial \\Internet of Things}

\author{\IEEEauthorblockN{Hao Wu, Hui Tian, \IEEEmembership{Senior Member, IEEE}, Shaoshuai Fan and Jiazhi Ren
}\vspace{-2em}

\thanks{H. Wu is with the State Key Laboratory of Networking and Switching Technology,
Beijing University of Posts and Telecommunications, Beijing 100876, China, and also with the Center for Complex Network Research, Northeastern University, Boston, MA 02115, USA. (e-mail: wh9405@bupt.edu.cn).}
\thanks{H. Tian, S. Fan, and J. Ren are with the State Key Laboratory of Networking and Switching Technology,
Beijing University of Posts and Telecommunications, Beijing 100876, China (e-mail: tianhui@bupt.edu.cn; fanss@bupt.edu.cn; renjiazhi6150@gmail.com).}

}

\maketitle

\begin{abstract}
Wireless powered mobile edge computing has been envisioned as a promising paradigm to enhance the computation capability of low-power wireless devices in Industrial Internet of Things.
An efficient resource scheduling method is critical yet challenging to design in such a scenario due to stochastic traffic arrival, time-coupling uplink/downlink decision and incomplete system state knowledge.
To tackle these challenges, an online optimization algorithm is proposed in this paper to maximize long-term system utility balancing throughput and fairness, subject to data age and stability constraints.
A set of virtual queues is designed to transform the scheduling task, which is hard to solve due to time-dependent data age constraints, into a stochastic optimization problem.
Leveraging Lyapunov and convex optimization techniques, the proposed approach can achieve asymptotically near-optimal online decisions without any prior statistical knowledge, and maintain the asymptotic optimality in the presence of partial and outdated network state information.
Numerical simulations corroborate the theoretical analysis and demonstrate the effectiveness of the proposed approach.
\end{abstract}

\begin{IEEEkeywords}
Wireless power transfer, mobile edge computing, Industrial Internet of Things, data freshness, stochastic optimization.
\end{IEEEkeywords}

\vspace*{-10pt}
\section{Introduction}
%
%
%
%

\IEEEPARstart{I}{ndustrial} Internet of Things (IIoT) is an emerging domain that promises ubiquitous interaction between the physical world and its digital counterpart~\cite{bibli:IIoT4}.
To this end, a large number of wireless devices (WDs) are placed in IIoT to enable ambient intelligence through continuous environmental monitoring and data analysis (e.g., hazardous gas detection in mining).
However, these services are usually data- and computation- intensive~\cite{bibli:IIoT3}, thus unfit for IIoT devices whose battery size and computation capacity are limited.

Wireless powered mobile edge computing (MEC) has been foreseen as a promising technology to tackle the above problems~\cite{bibli:delay2,bibli:WMEC1,bibli:delay3,bibli:deep,bibli:UAV,bibli:cooperation}.
Leveraging both the advantages of wireless power transfer (WPT)~\cite{bibli:UAV,bibli:cooperation,bibli:protocol2} and MEC~\cite{bibli:delay,bibli:channel,bibli:Ren,bibli:outdated} technologies, wireless powered MEC can promote device sustainability and provide powerful computation capability through efficient scheduling schemes.
However, scheduling in wireless powered MEC scenarios is more complex and tricky than that of separate WPT/MEC networks.
The reason mainly lies in two aspects.
On the one hand, network resources (e.g., WPT time portion and offloading bandwidth) in the scenario are highly coupled and dependent on each other.
For example, the delivery of wireless energy occupies the spectrum for data offloading.
On the other hand, to deal with the \emph{doubly near-far problem}~\cite{bibli:WMEC1} caused by the distance differences between the WPT node and its serviced WDs, the scheduling decision is usually made by an access point (AP, embedded with WPT and MEC functionalities) in a centralized manner.
The centralized decision would require frequent status feedback from WDs, which may degrade system performance when signaling overhead is non-negligible.
Although there exists some work to address the above scalability problem with partial feedback~\cite{bibli:channel,bibli:Ren,bibli:outdated}, few studies have been conducted for IIoT, where new application demands emerge.

Unlike conventional wireless powered MEC networks, data analysis services in IIoT have more requirements on data freshness~\cite{bibli:freshness}.
For instance, assembly line detecting service provides the status of products to stakeholders every twenty seconds.
Data that have been collected one minute ago contribute nothing to the analysis anymore.
To measure the freshness of data, a metric called data \emph{age} was defined in~\cite{bibli:drop}, which characterizes the time elapsed from being collected to being sent to a destination node.
It is worth noting that age here is quite different from the concept of \emph{delay} we usually encounter in scheduling.
A low collecting frequency leads to a short data queueing, thus little delay, but may result in a high age due to early admission~\cite{bibli:update}.
In~\cite{bibli:update}, Yin \emph{et al.} also found that data age first decreases and then increases with respect to collecting frequency, while delay maintains increase.
This suggests that admission control is significantly important in keeping data fresh.
To the best of our knowledge, there has been no adequate consideration of data age in wireless powered MEC scenarios.
Existing scheduling schemes are either delay-constrained~\cite{bibli:delay2,bibli:WMEC1}, or not constrained at all~\cite{bibli:delay3,bibli:deep,bibli:UAV,bibli:cooperation}, which are unable to be extended to IIoT networks.

This paper studies the resource scheduling problem of wireless powered MEC in IIoT networks, where the WPT and data offloading process share the same spectrum in a time division multiple access (TDMA) manner.
Distinct from existing works, the offloaded data are freshness-required (i.e., their age has to be kept beneath a certain threshold).
Our target is to maximize the long-term system utility, which consists of throughput, fairness and age-related data processing penalties, in the absence of real-time network state information (NSI).
The key contributions of this article are as follows.
\begin{enumerate}
  \item A data age aware scheduling mechanism of wireless powered MEC for IIoT is proposed to jointly optimize data collection, wireless power transfer, and data offloading.
      Lyapunov optimization is applied to decouple the system utility maximization problem over slots and among devices.
      Leveraging convex optimization techniques, we obtain the optimal solutions in closed and semi-closed forms and prove their asymptotic optimality.

  \item We design a set of age-aware virtual queues to decouple the time dependence brought by data age constraints.
      Apart from reducing computational complexity, the introduction and optimization of the age-aware virtual queues also guarantee the freshness of offloaded data.
  \item An analytic framework is proposed to reduce the demand for frequent feedback by using outdated partial NSI to estimate device states.
      The optimality loss caused by partial feedback and state estimation on long-term system utility is analyzed, and is proved to diminish as the control parameter of Lyapunov increases.
\end{enumerate}

Extensive simulations verify the asymptotic optimality of the proposed data age aware scheduling mechanism of wireless powered MEC in IIoT.
Compared with state-of-the-art schemes, our proposed approach improves system throughput, ensures fairness, and provides fresher data by dynamically adapting to varying network traffic and wireless channels.

It is worth noting that the adoption of the Lyapunov optimization technique in this article is to design an asymptotically optimal and practical scheduling approach under data age constraints.
Although many conventional stochastic optimization techniques have been studied in the literature, none of them could be applied to the considered IIoT scenario.
On the one hand, the data age constraints make the system optimization over different slot coupled and difficult to tackle.
On the other hand, the [$\mathcal{O}(V)\!-\!\mathcal{O}(1/V)$] tradeoff of standard Lyapunov optimization may result in great optimality loss to guarantee the data age constraints.
To deal with these challenges, in the proposed algorithm, we propose a $\epsilon$-persistent service queue based virtual queue technique, which transforms data age constraints into queue stability problems.
Convex optimization and Lambert W function are then leveraged to efficiently solve the transformed problems.

The rest of this paper is organized as follows.
In Section~II, we summarize some related work.
In Section~III, the system model is described.
In Section~IV, we present the data age aware scheduling policy that maximizes the time-average system utility, followed by the extrapolation to a more practical IIoT scenario with partial network knowledge.
In Section~V, the merit of the proposed scheduling mechanism is validated through simulations.
Finally, we conclude this paper in Section~VI.

\vspace*{-8pt}
\section{Related Work}
\vspace*{-2pt}
Recent years have witnessed encouraging progress on joint energy transfer and data offloading optimization of wireless powered MEC~\cite{bibli:delay2,bibli:delay3,bibli:deep,bibli:UAV,bibli:WMEC1,bibli:cooperation}.
In~\cite{bibli:delay2}, the optimization was formulated as an AP's energy consumption minimization problem under partial offloading, and solved by taking Lagrange duality methods.
Unlike~\cite{bibli:delay2}, reference~\cite{bibli:delay3} proposed an approach with a bi-section search and coordinate descent method to maximize the sum computation rate of multiple WDs that follows a binary offloading policy.
A similar problem was also considered in~\cite{bibli:deep} through deep reinforcement learning.
Inspired by these prior works, authors in~\cite{bibli:UAV} then extended their research to an unmanned aerial vehicle (UAV) scenario under both partial and binary offloading modes, where communication and computation resources, as well as the trajectory of UAV are jointly designed to maximize achievable computation rate.
In~\cite{bibli:WMEC1} and~\cite{bibli:cooperation}, cooperative resource allocation among WDs was investigated from the perspectives of maximizing throughput~\cite{bibli:WMEC1} and energy efficiency~\cite{bibli:cooperation}.
However, all these schemes require the full knowledge of real-time network states, whereas in practice, only partial and outdated NSI is available due to non-negligible feedback overhead and signaling transmission delay.
Hence, they are not directly applicable to IIoT where the traffic arrival and computation processing rate are time-varying.

Stochastic optimization, such as stochastic gradient descent~\cite{bibli:SGD,bibli:timescale,bibli:SGD2} and standard Lyapunov optimization~\cite{bibli:channel,bibli:Ren,bibli:outdated,bibli:myIoT}, is the underlying technique to provide a satisfactory performance in the absence of full network knowledge.
The idea of this approach is to leverage its asymptotic nature to diminish the optimality loss brought by using outdated NSI as an approximation of current system states.
Moreover, stochastic optimization approaches can also decouple the optimization of a stochastic system over slots, which prevent the system from running into the \emph{curse-of-dimensionality problem} when network variables are a huge number~\cite{bibli:delay}.
To the best of our knowledge, an online scheduling mechanism with data age awareness has yet to be developed for wireless powered MEC.
The most relevant work is~\cite{bibli:myIoT}, which uses a similar Lyapunov technique for scheduling optimization.
However, it did not consider any data freshness requirement in IIoT.
Moreover, algorithms developed there still need devices to send back their device state information every time they get the opportunity to offload.
Hence, the problem considered in this paper is distinctively different from the typical Lyapunov technique adopted in~\cite{bibli:myIoT}.

\vspace*{-7pt}
\section{System Model}
\subsection{System Overview}
\begin{figure}
\centering{\includegraphics[height=0.45\textwidth]{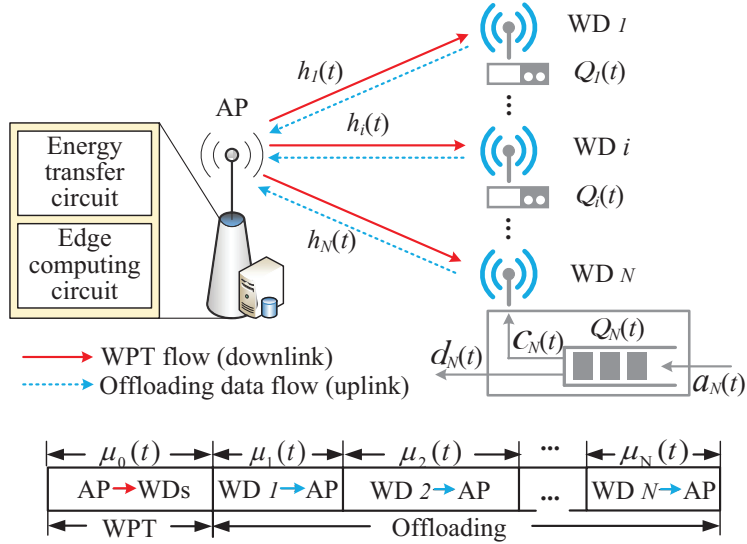}}
\caption{A wireless powered MEC network model in IIoT with data age awareness.}
\label{fig:scenario}
\end{figure}

The network of interest consists of an AP and a set of $N$ WDs indexed by $\mathcal{N} \!=\! \{ 1,\cdot \cdot \cdot,N\}$.
Similar to most works in wireless powered MEC~\cite{bibli:WMEC1,bibli:delay3,bibli:deep,bibli:UAV,bibli:cooperation,bibli:myIoT}, WPT and data offloading are operated over the same frequency band in a frame-based TDMA mode.
For convenience, the time duration of every frame is normalized to a unit length in the sequel.
As illustrated in Fig.~\ref{fig:scenario}, each frame is divided into two phases, i.e., WPT phase and offloading phase.
In the WPT phase, WDs simultaneously harvest wireless energy from the AP during the allocated $\mu_0(t)$ time portion.
Then in the remaining $1\!-\!\mu_0(t)$ time, WDs take turns to offload their data to the AP by using the energy harvested in the WPT phase.
The time portion assigned for WD $i\! \in \!\mathcal{N}$ to offload is denoted by $\mu_i(t)$.
Given data analysis services for IIoT, the computing results at the AP do not need to be sent back to WDs~\cite{bibli:myIoT} or are neglected due to their small size compared with offloaded data~\cite{bibli:delay3,bibli:deep,bibli:UAV,bibli:WMEC1,bibli:cooperation}.
Therefore, the time portions $\tilde{\bm{\mu}}(t)\!=\!(\mu_i(t),i\in\mathcal{\tilde{N}})$ in each time slot satisfy
\begin{equation}
\sum\nolimits_{i \in \mathcal{\tilde{N}}} \mu_i(t) \leq 1,~\mu_i(t)\geq 0,~\forall i \in \mathcal{\tilde{N}},
\vspace*{-3pt}
\label{equ:mu}
\end{equation}
where $\mathcal{\tilde{N}} \triangleq \{0\}\cup\mathcal{N}$ represents the set of the AP and $N$ WDs.
It is worth mentioning that even if the time for downloading the computing results cannot be neglected and accounts for a proportion $\alpha$ of each time slot, the proposed algorithm in this paper can be easily extended.
If $\alpha$ is prior known constant, we can extend the algorithm by directly scaling the normalized unit slot to $1\!+\!\alpha$.
Otherwise, when $\alpha$ changes from slot to slot, the result downloading could be seen as an inverse process of uplink network state information feedback, which has been well-studied in~\cite{bibli:myIoT}.
At every slot, we can first determine the set of WDs that need downloading, excluding their downlink time portions from the unit slot duration based on result sizes, and then apply the proposed approach for joint WPT and data offloading optimization.

Both the downlink (from the AP to WDs) and uplink (from a WD to the AP) channels are assumed to be independent and identically distributed (i.i.d.) flat block fading~\cite{bibli:delay,bibli:channel}.
In other words, channels remain static within a single time slot, but may vary with respect to different slots.
The wireless channel gain between the AP and WD $i$ is denoted as $h_i(t)$.
Consider the channel reciprocity of TDMA mode system~\cite{bibli:protocol2,bibli:delay3}, $h_i(t)$ is reciprocal for the downlink and uplink.
At the beginning of each time slot, the AP acquires $h_i(t)$ by receiving the pilot for channel measurement from each WD.
Due to the small amount of transmission, the signaling cost and energy consumption for estimating $h_i(t)$ are neglected~\cite{bibli:delay2,bibli:WMEC1,bibli:delay3,bibli:deep,bibli:UAV,bibli:cooperation}.

In the considered system, we assume the AP to be equipped with a single antenna and a reliable power supply.
The constant transmit power $P_0$ of the AP is relatively large such that the energy harvested from noises and received uplink signals from other WDs can be neglected~\cite{bibli:protocol2}.
Recall that the duration of each slot is normalized to be unit, the energy harvested by WD~$i$ in the WPT phase of slot $t$ can be written as~\cite{bibli:delay3,bibli:myIoT}
\begin{equation}
E_i(t) = \xi_i  P_0 h_i(t) \mu_0(t), \ \  i \in \mathcal{N},
\vspace*{-3pt}
\label{equ:energy}
\vspace*{-3pt}
\end{equation}
where $\xi_i \in (0,1)$ is the energy harvesting efficiency.

Given the manufacturing complexity and costs, IIoT WDs for sensing applications are usually equipped with rechargeable batteries that have high self-discharge rate and low energy storage capacity~\cite{bibli:protocol2,bibli:myIoT}.
In this case, they can only hold the harvested energy for a slot duration and are inclined to exhaust all the harvested energy for offloading within each time slot~\cite{bibli:delay3,bibli:deep,bibli:UAV}.
As the harvested energy is not a constant quantity and WDs' offloading time varies with time, the transmit power of WDs are varying from slot to slot, the dynamic coordination of which can be supported by embedded WPT printed circuit boards~\cite{bibli:sensor,bibli:3GPPmy}.
Hence, the transmit power of any WD $i\in \mathcal{N}$ (i.e., $P_i(t)$) equals to $E_i(t)/\mu_i(t)$.
Combining this with equation~(\ref{equ:energy}), the achievable offloading rate of WD $i$ within slot $t$ can be expressed as
\begin{equation}
\begin{aligned}
c_i(t) &\!=\! \mu_i(t) W \log_{2}(1+\frac{P_i(t)h_i(t)}{N_0}  )\\
&\!=\! \mu_i(t) W \log_{2}(1+\frac{\delta_i(t)\mu_0(t)}{\mu_i(t)}  ), \ \  i \in \mathcal{N},
\end{aligned}
\label{equ:capacity1}
\end{equation}
where $W$ denotes the communication bandwidth, $N_0$ represents the noise power at the receiver of the AP, and $\delta_i(t)\!=\!\xi_i P_0 h_i^2(t)/N_0$ is a constant in an individual slot.
The maximum transmit power constraint of WDs is not considered since the energy harvested from WPT are of small amount in practice~\cite{bibli:delay3,bibli:protocol2}.
Given the small uplink transmission power, $c_i(t)$ in (\ref{equ:capacity1}) is upper bound by a maximum link capacity $c_i^{\text{max}}$.
The notations used in this paper are summarized in Table I.

\newcommand{\tabincell}[2]{
}
\begin{table}[!t]
  \centering
  \scriptsize
  \caption{Summary of Notations}
  \label{tab:notations}
  \begin{tabular}{ll}
    \\[-2mm]
    \hline
    \hline\\[-2mm]
    {\bf \small Notation}& {\bf\small Description}\\
    \hline
    \vspace{1mm}\\[-3mm]
    $\mu_0(t)$  &  Time portion for wireless power transfer at slot $t$\\
    \vspace{1mm}
    $\mu_i(t)$  &  Time portion for device $i$ to offload at slot $t$\\
     \vspace{1mm}
    $E_i(t)$  &   Harvested energy of device $i$ at slot $t$\\
     \vspace{1mm}
    $\xi_i$   &  Energy harvesting efficiency of device $i$\\
    \vspace{1mm}
    $P_0$   &  Transmit power of the AP \\
    \vspace{1mm}
    $P_i(t)$   &  Offloading transmit power of device $i$ at slot t\\
    \vspace{1mm}
    $h_i(t)$   &  Channel power gain between AP and device $i$ at slot t \\
    \vspace{1mm}
    $W$      &  System bandwidth \\
    \vspace{1mm}
    $c_i(t)$    & Achievable offloading data size of device $i$ at slot $t$ \\
    \vspace{1mm}
    $A_i(t)$  &   Data that can be collected by device $i$ within slot $t$\\
     \vspace{1mm}
    $a_i(t)$  &   Data collected at device $i$ within slot $t$\\
    \vspace{1mm}
    $d_i(t)$  &   Data dropped at device $i$ within slot $t$\\
    \vspace{1mm}
    $r_i(t)$  &   Available data processing speed of AP at slot $t$\\
    \vspace{1mm}
    $Q_i(t)$  &   Data queue backlog of device $i$ at slot $t$\\
    \vspace{1mm}
    $S_i(t)$  &   Data queue backlog of AP at slot $t$\\
    \vspace{1mm}
    $Z_i(t)$  &   Virtual queue backlog of AP at slot $t$\\
    \vspace{1mm}
    $g_{i,t_j}(t)$  &   Age of data collected at slot $t_j$ by device $i$ during slot $t$\\
    \vspace{1mm}
    $\overline{X}$  &   Time-average of any stochastic process $X$\\
     \vspace{1mm}
    $\phi_i(\cdot)$  &  System benefit brought by data collection from device $i$\\
    \hline
    \hline
  \end{tabular}
\end{table}

\subsection{Data Collection, Queueing, and Processing}
At every time slot $t$, the available volume of data that could be collected by $i$-$th$ WD is denoted by $A_i(t)$.
Due to the ever-changing nature of IIoT system, $A_i(t)$ is varying over slots and is modeled as i.i.d. with upper bound $A_i^{\text{max}}$~[11], [20].
Given the limited data buffer and data freshness of WDs, only part of the $A_i(t)$ amount of data, denoted as $a_i(t)$, can be collected into $i$-$th$ WD's data buffer.
Therefore, $a_i(t)$ satisfies
\begin{equation}
0 \leq a_i(t) \leq A_i(t)\leq A_i^{\text{max}}, ~\forall i \in \mathcal{N}.
\label{equ:a}
\end{equation}
Given the time cost for data collection, the amount of data $a_i(t)$ will become processable from the beginning of slot~$t\!+\!1$.
Before there are available wireless uplink channels, collected data are all stored in the data buffer maintained by its WD.

The congested wireless channels may result in a long waiting time of data queued in the WD's buffer.
For age-sensitive data services, stale data whose age exceed a certain threshold make no contribution.
Therefore, to provide memory for fresh data and relieve the burden of wireless channels, WDs would drop the data that are about to violate the freshness requirement~\cite{bibli:drop}.
Given the maximum data a WD $i$ can collect in one slot, the amount of discarded data $d_i(t)$ is subject to
\begin{equation}
0 \leq d_i(t) \leq A_i^{\text{max}},~\forall i \in \mathcal{N}.
\label{equ:d}
\end{equation}

Let $\bm{Q}(t)\!=\!(Q_i(t),i \in \mathcal{N})$ be the data queue backlog vector of WDs in the network.
At every slot, the system decides the amount of data to be collected (i.e., $a_i(t)$), can be offload (i.e., $c_i(t)$), and need to be dropped (i.e., $d_i(t)$) based on network states.
Follow the basic first-in-first-out (FIFO) model of cascaded queueing systems~\cite{bibli:channel,bibli:myIoT}, the queueing dynamics of the data backlog at WD~$i$ is given by
\begin{equation}
Q_i(t+1) = [Q_i(t) - c_i(t) - d_i(t)]^+ + a_i(t),
\label{equ:Qt}
\end{equation}
where $[x]^+ \!=\! \max(x,0)$.
The first term on the right-hand-side (RHS) of (\ref{equ:Qt}) accounts for the data remained at the end of slot $t$, after part of the data has been offloaded or dropped.
When $Q_i(t) \leq c_i(t)\!+\!d_i(t)$, the system will give priority to data offloading, and then discard the remaining data.

At the AP side, $N$ data buffers $\bm{S}(t)\!=\!(S_i(t),i \in \mathcal{N})$ are maintained to store the data offloaded from WDs but not yet executed by the AP.
Given the limited computing capability of the AP, at most $r_i(t)$ amount of data from WD $i$ can be processed in slot $t$, where $r_i(t)$ is a stochastic number with the with maximum $r_i^{\text{max}}$~\cite{bibli:channel,bibli:myIoT}.
Therefore, the data backlog $S_i(t)$ evolves as
\begin{equation}
S_i(t+1) = [S_i(t) - r_{i}(t)]^+ + \min\{c_i(t),Q_i(t)\},
\label{equ:Ft}
\end{equation}
where the second term on RHS specifies that the actual offloaded data could not outnumber the amount of data in $i$-th WD's buffer, when the allocated $c_i(t)$ is larger than $Q_i(t)$.

\section{Data Age Aware Online Optimization of Wireless Powered MEC}
We first present a new asymptotically optimal online approach with data age awareness for data update (i.e., collection and discard), WPT, and data offloading, provided instantaneous network state knowledge is available.
Then we will prove that the instantaneous optimization can be readily extended to the scenario with partial and outdated NSI, while preserving the asymptotic property.

\subsection{Problem Formulation and Algorithm Design}
To get better overall system performance, data analysis services in IIoT pay attention to not only throughput but also fairness among WDs~\cite{bibli:myIoT}.
However, fairness is not a easy task in wireless powered MEC networks due to the doubly near-far problem.
To tackle this issue, we define the instantaneous system utility at any time slot $t$ as
\begin{equation}
U(\bm{v}^t)=\sum\nolimits_{i \in \mathcal{N}}\big[~\phi_i\big(a_i(t)\big)-p_i(t)d_i(t)~\big],
\label{equ:U}
\end{equation}
where $\bm{v}^t\!=\!\{\mu_0(t);\mu_i(t),a_i(t),d_i(t)\!:\!\forall i\in \mathcal{N}\}$ collects all variables to be optimized at slot $t$, $\phi_i(a_i(t))$ denotes the system benefit brought by data collection, and $p_i(t)$ is the price of WD~$i$ for dropping one bit of data.
Given system fairness, similar to~\cite{bibli:myIoT}, we choose $\phi_i(x)\!=\!\log(1\!+\!x)$, the decreasing marginal utility of which encourages the equal collection of data from different WDs.
Based on proportional fairness~\cite{bibli:channel,bibli:myIoT}, the same effect could also be achieved as long as $\phi_i(x)$ is continuous, concave and non-decreasing with $\phi_i(0)\!=\!0$.
We also assume that $p_i(t)\!=\!p$, where $p\! \geq \!1$, for arbitrary WD $i$ and slot $t$, since no WD priority is considered in this paper.
As can be seen from (\ref{equ:U}), there is an explicit tradeoff between data collection and discard.
A frequent data collection leads to the increase of the first term on the RHS, but may cause the ageing and dropping of queueing data, which results in the degradation of $U(\bm{v}^t)$ when $p$ is high.

Denote $g_{i,t_j}(t)$ as the age of data during slot $t$ that was collected by WD $i$ at time slot $t_j$.
Given the short duration of each time slot and for analytical tractability, it is assumed that data age within any slot $t$ will not be impacted by the offloading sequence of WDs and remains unchanged in one slot.
Therefore, according to the definition of data age (i.e., the time elapsed since the data was collected~\cite{bibli:drop}), we have $g_{i,t_j}(t)\!=\!t\!-\!t_j$.
Data queued at a WD is deemed fresh and valuable if
\begin{equation}
g_{i,t_j}(t) \leq g_{\text{max}},~\forall i \in \mathcal{N},
\label{equ:age}
\end{equation}
where $g_{\text{max}}$ is the maximum acceptable age of the IIoT system for data analysis.

Consider the long-term time-average performance of the system, the problem of interest can be formulated as
\begin{equation*}
\begin{aligned}
\textbf{P1}:~&\mathop{\max}_{\{\bm{v}^t,\forall t\}}~\overline{U(\bm{v}^t)}
\\ &~~~~~~\text{s.t.} ~~(\ref{equ:mu}),(\ref{equ:a}),(\ref{equ:d}),(\ref{equ:age}),\\
&~~~~~~~~~~~\textbf{C1:}~\overline{Q_i}< \infty,~\overline{S_i}< \infty,~\forall i \in \mathcal{N},
\end{aligned}
\label{equ:utility0}
\end{equation*}
where $\overline{X}\!=\!\lim_{t \to \infty} (1/t) \sum\nolimits_{\tau =0}^{t-1} \mathbb{E}\{ X(\tau) \}$ defines the time average expectation of any stochastic process $X(t)$, and constraint \textbf{C1} ensures the stability of all data queues in the network~\cite{bibli:Lyapunov}.

In wireless powered MEC systems, the burst data arriving at the AP may take up all the computational resources and cause queueing backlogs.
A typical example in IIoT is warehouse environmental monitoring services, where the AP, out of manufacturing cost, is designed with finite computing capacity to deal with most of the daily cases instead of peak requirements.
Therefore, we coordinate the data to be offloaded to match the available computation capability the AP has at different time slots through the constraint \textbf{C1}.
Although the considered problem aims at sending data to the AP with throughput as one of its metrics, it is fundamentally different from uplink data transmission in wireless powered communication networks (WPCNs)~\cite{bibli:protocol2}, where the AP is equipped with unlimited computation capability such that data arrived at the AP can be processed immediately.

The problem above can in principle be solved by approaches such as game-theoretic, heuristic, or reinforcement learning-based methods.
However, they may require non-causal NSI (e.g.,game-theory), lead to myopic schedule with great optimality loss (e.g., heuristic method), or result in high complexity and learning time (e.g., deep Q-learning).
Although stochastic optimization is a good choice to tackle this type of long-term problem, it cannot be directly applied to problem \textbf{P1} since (\ref{equ:age}) is time-dependent and coupled with the decision of $a_i(t)$ and $d_i(t)$, which leads to the dependency of $a_i(t)$ and $d_i(t)$ on the past.
The past values of $a_i(\tau)$ and $d_i(\tau)$ for $\tau\!<\!t$ would influence the current queue states $\bm{Q}(t)$ and $\bm{S}(t)$, resulting in the skew of the conditional distribution of $a_i(t)$ and $d_i(t)$.

To address the time-coupling problem caused by~(\ref{equ:age}), we meticulously design a set of virtual queues $\bm{Z}(t)\!=\!(Z_i(t),i \in \mathcal{N})$ for every WD, which evolves as
\begin{equation}
Z_i(t+1)\!=\![Z_i(t)-\frac{1}{p}c_i(t)-p d_i(t)+p\epsilon_i]^+,
\label{equ:Zt}
\end{equation}
where $\epsilon_i\! \in \!(0,A_i^{\text{max}}]$ is a constant value indicates the arrival rate of $Z_i(t)$.
We will later prove in Section~IV-C that if the stability of $Z_i(t)$ can be ensured, i.e.,
\begin{equation}
\overline{Z_i}< \infty,~\forall i \in \mathcal{N},
\label{equ:Zstable}
\end{equation}
the size of the queue $Z_i(t)$ can provide a bound on the data age of the head-of-line data in $Q_i(t)$.
If we can adjust the upper bound of data age in each WD, denoted as $g_i^{\text{max}}$, to be no larger than $g_{\text{max}}$, the age constraint (\ref{equ:age}) can always be guaranteed.
Therefore, by replacing~(\ref{equ:age}) with~(\ref{equ:Zstable}), the problem \textbf{P1} can be equivalently transformed to a typical stochastic optimization problem.
With the aid of Lyapunov optimization technique~\cite{bibli:Lyapunov}, such a problem can be further decoupled between time slots in an asymptotically optimal manner, as stated in the following theorem.

\renewcommand{\IEEEQED}{\IEEEQEDopen}
\newtheorem{lemma}{Lemma}
\begin{lemma}
The problem \textbf{P1} can be reformulated to a deterministic per-slot problem as follows
\begin{equation}
\begin{aligned}
\textbf{P2}:~&\mathop{\min}_{\bm{v}^t}~f_1(\bm{a}(t))+f_2(\bm{d}(t))+f_3(\tilde{\bm{\mu}}(t))
\\ &~~~~~~\text{s.t.} ~~(\ref{equ:mu}),(\ref{equ:a}),(\ref{equ:d}),
\end{aligned}
\vspace*{-10pt}
\label{equ:P3}
\end{equation}
where
\begin{subequations}
\begin{align}
&f_1(\bm{a}(t))=\sum\nolimits_{i \in \mathcal{N}}\left[~Q_i(t)a_i(t)-V\phi_i\big(a_i(t)\big)~\right],\\
&f_2(\bm{d}(t))=\sum\nolimits_{i \in \mathcal{N}}\left[~Vp-Q_i(t)- Z_i(t)~\right]d_i(t),\\
&f_3(\tilde{\bm{\mu}}(t))\!=\!\sum\nolimits_{i \in \mathcal{N}}\big[S_i(t)\!-\!Q_i(t)\!-\!\frac{1}{p}Z_i(t)\big] c_i(t),
\end{align}
\end{subequations}
are the decision functions on data collection, data discard, and offloading rate at time slot $t$, respectively; and $V$ is a nonnegative control parameter that affects the optimality bound of overall system utility and $g_i^{\text{max}}$, as will be stated in Theorem 1 and 2.
Based on different data freshness requirements of system devices, the value of $V$ is adjustable.
As long as (\ref{equ:age}) can be guaranteed, selecting a sufficiently large $V$ can diminish the optimality loss brought by the asymptotical optimization.
\label{lemma:problem}
\end{lemma}
\begin{IEEEproof}
Please refer to Appendix A
\end{IEEEproof}

\newtheorem{remark}{Remark}
Note that $\bm{a}(t)=(a_i(t),i \in \mathcal{N})$, $\bm{d}(t)=(d_i(t),i \in \mathcal{N})$, $\tilde{\bm{\mu}}(t)=(\mu_i(t),i \in \mathcal{\tilde{N}})$ can be decoupled from each other in both the objective and constraints.
Besides, $a_i(t)$ and $d_i(t)$ can be decoupled among different WDs since they are independently determined by each device.
Accordingly, the minimization problem \textbf{P2} can be decomposed into three sub-problems, namely, fresh data collection, stale data discard, and offloading decision as follows

\begin{equation}
    \mathop{\min}_{a_i(t)}~Q_i(t)a_i(t)-V\phi_i\big(a_i(t)\big),~~\text{s.t.~(\ref{equ:a})},
    \label{equ:collection}
    \vspace*{-5pt}
\end{equation}

\begin{equation}
    \mathop{\min}_{d_i(t)}~\left[~Vp-Q_i(t)- Z_i(t)~\right]d_i(t),~~\text{s.t.~(\ref{equ:d})},
    \label{equ:discard}
    \vspace*{-5pt}
\end{equation}

\begin{equation}
   \mathop{\min}_{\tilde{\bm{\mu}}(t)}~f_3(\tilde{\bm{\mu}}(t)),~~\text{s.t.~(\ref{equ:mu})},
   \label{equ:powerdata}
   \vspace*{-5pt}
\end{equation}
which minimizes the unnecessary cost for redundancy data collection, keeps the freshness of queued data at devices, and maximizes the amount of offloading data.

\begin{algorithm}[t]
\caption{Data Age Aware Online Optimization of Wireless Powered MEC}
\label{alg:1}
\LinesNumbered 
\textbf{For the AP at every time slot $t$}:\\
1:~Observe $\bm{Q}(t)$, $\bm{S}(t)$, $\bm{Z}(t)$ and channel gains;\\
2:~Optimize $\tilde{\bm{\mu}}(t)$ by solving (\ref{equ:powerdata});\\
3:~Inform $\tilde{\bm{\mu}}(t)$ to WDs and execute WPT;\\
\textbf{For any WD $i$ at time slot $t$}:\\
4:~Observe $Q_i(t)$, $Z_i(t)$, $A_i(t)$ and $c_i(t)$;\\
5:~Decide $a_i(t)$ by solving (\ref{equ:collection});\\
6:~Choose $d_i(t)$ by solving (\ref{equ:discard});\\
7:~Offload data and give feedback $Q_i(t)$ and $Z_i(t)$ to the AP in the arranged sequence;\\
8:~Update $Q_i(t)$ according to (\ref{equ:Qt})\\
9:~Update $Z_i(t)$ according to~(\ref{equ:Zt})\\
\textbf{For the AP at every time slot $t$}:\\
10:~Update $\bm{S}(t)$ according to (\ref{equ:Ft}).
\end{algorithm}

Algorithm 1 summarizes the proposed data age aware online optimization of wireless powered MEC.
It is worth noting that the decisions of data collection and discard can be made by WDs themselves in a distributed manner, while the optimization of $\tilde{\bm{\mu}}(t)$ has to be done in a centralized way by the AP due to the device-dependence brought by constraint~(\ref{equ:mu}).
Next, we develop efficient solvers of (\ref{equ:collection}), (\ref{equ:discard}), and (\ref{equ:powerdata}) to facilitate the online scheduling.

\subsection{Per-slot Optimal Solutions of Online Optimization}
Problems (\ref{equ:collection}) and (\ref{equ:discard}) are both convex optimization problems, since their objectives and constraints are all convex.
Therefore, their optimums are either achieved at their stationary points or one of the boundaries, which are given by
\begin{equation}
a_i^{*}(t)=
\begin{cases}
A_i(t),& \text{$V \geq \big(A_i(t)+1\big)Q_i(t)$};\\
[\frac{V}{Q_i(t)}-1]^+,& \text{otherwise};
\end{cases}
\label{equ:condition1}
\end{equation}
\begin{equation}
d_i^*(t)=
\begin{cases}
A_i^{\text{max}},& \text{$Q_i(t)+ Z_i(t)>Vp$};\\
0,& \text{otherwise},
\end{cases}
\label{equ:drops}
\end{equation}
The given results above are in accordance with our intuition that when the data backlog at WD $i$ (i.e., $Q_i(t)$) is large, the device is inclined to collect less new data and drop more stale one (i.e., choosing smaller $a_i(t)$ and larger $d_i(t)$).

Problem (\ref{equ:powerdata}) is a non-convex problem in general.
Fortunately, we observe an important property that the optimal uplink time allocation of $i$-th WD is give by $\mu_i(t)=0$ for any $i\! \in \!\mathcal{N}_t\!=\!\{i | S_i(t)\!-\!Q_i(t)\!-\!\frac{1}{p}Z_i(t)\!\geq \!0\}$.
The proof of this property is simple since the term $[S_i(t)\!-\!Q_i(t)\!-\!\frac{1}{p}Z_i(t)]c_i(t)$ gets its minimum at $\mu_i(t)\!=\!0$ for arbitrary $i\! \in \!\mathcal{N}_t$.
The originally allocated time portions for them can thus be re-assigned to WPT and data offloading of WDs in $\mathcal{N}\backslash \mathcal{N}_t$ to achieve optimum system utility.
Such results are also consistent with our intuition that when there are too much data waiting to be processed at the AP, it is better to reduce offloaded data to avoid unnecessary system overhead and execution delay.
By excluding the WDs in $\mathcal{N}_t$, problem~(\ref{equ:powerdata}) can then be rewritten as
\begin{equation}
\begin{aligned}
\mathop{\min}_{\tilde{\bm{\mu}}_c(t)}~~W&\sum\nolimits_{i \in \mathcal{N}\backslash \mathcal{N}_t}\big[S_i(t)-Q_i(t)\\
&~~~~~~-\frac{1}{p}Z_i(t)\big] \mu_i(t) \log_{2}(1+\frac{\delta_i(t)\mu_0(t)}{\mu_i(t)})
\\ \text{s.t.} ~~\mu_0(t)\!+\!&\sum\nolimits_{i \in \mathcal{N}\backslash \mathcal{N}_t} \mu_i(t) \leq 1,~\mu_i(t)> 0,~\forall i \in \mathcal{\tilde{N}}\backslash \mathcal{N}_t,
\end{aligned}
\label{equ:exclu}
\end{equation}
where $\tilde{\bm{\mu}}_c(t)\!=\!\{\mu_i(t) | i\! \in \! \mathcal{\tilde{N}} \backslash \mathcal{N}_t \}$.

As the objective of (\ref{equ:exclu}) is a non-positive sum of the perspective function of the concave function $f(x)\!=\!\log_{2}(1\!+\!\delta_i(t) x)$, it is convex in \{$\mu_0(t),\mu_i(t)\!:\!\forall i\in \mathcal{N}\backslash \mathcal{N}_t$\}~\cite{bibli:convex}.
It readily follows that (\ref{equ:exclu}) is a convex optimization problem, which can be effectively solved by Karush-Kuhn-Tucker (KKT) optimization~\cite{bibli:convex}.
\begin{lemma}
\label{lemma:mu}
Denote $\lambda^*(t)$ as the optimal Lagrange multiplier of KKT conditions at slot $t$. The optimal solutions to (\ref{equ:exclu}) satisfy
\begin{equation}
\mu_0^{*}(t) =
\frac{1}{1+\sum\nolimits_{i\in \mathcal{N}\backslash \mathcal{N}_t} \delta_i(t)\Phi_i(\lambda^{*}(t))},
\label{equ:mu0}
\vspace*{-5pt}
\end{equation}

\begin{equation}
\mu_i^{*}(t) = \frac{\delta_i(t)\Phi_i(\lambda^{*}(t))}{1+\sum\nolimits_{i\in \mathcal{N}\backslash \mathcal{N}_t} \delta_i(t)\Phi_i(\lambda^{*}(t))},~i\in \mathcal{N}\backslash \mathcal{N}_t,
\label{equ:mui}
\vspace*{-5pt}
\end{equation}
where
\begin{equation}
\Phi_i(\lambda^{*}(t))= -\left[ 1\!+\!\frac{1}{W_0\big(-\text{exp}({\frac{\ln2\lambda^{*}(t)}{S_i(t)\!-\!Q_i(t)\!-\!\frac{1}{p}Z_i(t)}-1})\big)}  \right]^{-1},
\label{equ:W}
\end{equation}
in which $W_0(x)$ is the principal branch of the Lambert W function defined as the solution for $W_0(x)\text{exp}(W_0(x))\!=\!x$~\cite{bibli:WMEC1}.
\end{lemma}
\begin{IEEEproof}
Please refer to Appendix B
\end{IEEEproof}

Given the monotonicity of $\Phi_i(\lambda^{*}(t)$, $\lambda^*(t)$ and the optimal time portion allocation can be obtained through a two-tier bi-section search with complexity $|\mathcal{\tilde{N}}\backslash \mathcal{N}_t|\log(\frac{1}{\kappa})\log(\frac{1}{\sigma})$~\cite{bibli:convex}, where $\kappa$ and $\sigma$ are the accuracy requirements for calculating (\ref{equ:W}) and $\lambda^*(t)$, respectively.

\subsection{Asymptotic Optimality under Instantaneous Network State Information}
As stated in Section IV-A, the centralized scheduling at the AP would require instantaneous network state feedback (i.e., $\bm{Q}(t)$ and $\bm{Z}(t)$) from WDs.
Let $U^\star$ denote the offline optimum of problem \textbf{P1} obtained by using non-causal NSI, and $U^*(\bm{v}^t)$ be the long-term average utility achieved by our proposed algorithm with instantaneous NSI.
The asymptotic optimality of $U^*(\bm{v}^t)$ can be proved by the following theorem.
\newtheorem{theorem}{Theorem}
\begin{theorem}
\label{lemma:O1V}
Suppose all queues are initially empty, the gap between $U^*(\bm{v}^t)$ and  $U^\star$ satisfies
\begin{equation}
U^\star - U^*(\bm{v}^t) \leq B_1/V,
\label{equ:gap1}
\end{equation}
where $B_1\!=\!\frac{1}{2}\max\{\epsilon_i^2,(A_i^{\text{max}}\!+\!\frac{1}{p^2}c_i^{\text{max}}\!-\!\epsilon_i)^2\}
\!+\!\frac{1}{2}[(c_i^{\text{max}}\!+\!A_i^{\text{max}})^2\!+\!(A_i^{\text{max}})^2\!+\!(c_i^{\text{max}})^2\!+\!(r_i^{\text{max}})^2]$ is a constant obtained from the proof of Lemma~\ref{lemma:problem}.
\end{theorem}
\begin{IEEEproof}
Please refer to Appendix C.
\end{IEEEproof}

\begin{theorem}
\label{lemma:OV}
The backlogs of $Q_i(t)$, $Z_i(t)$, and $S_i(t)$ are upper bounded by $Q_i^{\text{max}}$, $Z_i^{\text{max}}$, and $S_i^{\text{max}}$, respectively, which satisfy
\begin{subequations}
\begin{align}
Q_i^{\text{max}} &= V(2-e^{-p})+A_i^{\text{max}}; \label{equ:Qmax}\\
Z_i^{\text{max}} &= p(V + \epsilon_i); \label{equ:Zmax}\\
S_i^{\text{max}} &= Q_i^{\text{max}}+\frac{1}{p}Z_i^{\text{max}}+c_i^{\text{max}}. \label{equ:Smax}
\end{align}
\end{subequations}
The above results then guarantee that data age is upper bounded by
\begin{equation}
g_{i}^{\text{max}} = \lceil \frac{1}{\epsilon_i}(Q_i^{\text{max}} + \frac{1}{p}Z_i^{\text{max}}) \rceil,
\label{equ:gmax}
\end{equation}
\end{theorem}
where symbol $\lceil \cdot \rceil$ means rounding up to the nearest integer such that data age is an exact multiple of time slot duration.
\begin{IEEEproof}
Please refer to Appendix D
\end{IEEEproof}

Theorems \ref{lemma:O1V} and \ref{lemma:OV} reveal an explicit $[\mathcal{O}(1/V),\mathcal{O}(V)]$ tradeoff between system utility and data age.
As long as the age requirement is met, i.e., $g_i^{\text{max}}\leq g_{\text{max}}$ for any $i \in \mathcal{N}$, choosing a sufficiently large $V$ can diminish the optimality loss brought by the proposed algorithm and achieve close-to-optimal system utility.
In some IIoT applications where high reliability is required and no data is allowed to be dropped, the proposed approach can still be applied if we set $p\! \to\! \infty$.
According to (\ref{equ:drops}) and (\ref{equ:gmax}), even if $p$ is infinite, the proposed approach can guarantee a finite upper bound of data age by selecting an appropriate $V$.

\subsection{Asymptotic Optimality under Partial and Outdated Network State Information}

In IIoT networks, it is not practical to assume the instantaneous NSI is available since frequent feedback from WDs consumes a large number of spectrum resources.
In many cases, feedback is obtainable every $\tau\! \in\! \Omega \!=\!\{1,\cdots,m\}$ slots~\cite{bibli:timescale}, where $m$ is the maximum NSI feedback interval.
For this reason, we proceed to derive the solutions of Algorithm~1 by approximating the latest feedback values of queue backlogs as the current counterparts, which are given by
\begin{equation}
\widehat{Q}_i(t)= Q_i(t\!-\!\tau_i^Q),
\label{equ:Qa}
\vspace*{-3pt}
\end{equation}
\begin{equation}
\widehat{Z}_i(t) = Z_i(t\!-\!\tau_i^{Z}),
\label{equ:Za}
\vspace*{-3pt}
\end{equation}
where $\widehat{Q}_i(t)$ and $\widehat{Z}_i(t)$ denote the approximate backlogs of data and virtual queues at time slot $t$, respectively, $\tau_i^Q$ and $\tau_i^Z$ are the corresponding outdated time slots that belong to~$\Omega$.

As stated in Section IV, (\ref{equ:collection}) and (\ref{equ:discard}) are solved in a distributed manner by WDs themselves.
Therefore, the approximation only affects the optimization of (\ref{equ:powerdata}), which can be transformed to minimize
\begin{equation}
\begin{aligned}
\widehat{f_3}(\widehat{\tilde{\bm{\mu}}}(t))&=\!W\sum\nolimits_{i \in \mathcal{N}}[S_i(t)-\widehat{Q}_i(t)\\
&-\frac{1}{p}\widehat{Z}_i(t)] \widehat{\mu}_i(t) \log_{2}(1+\frac{\delta_i(t)\widehat{\mu}_0(t)}{\widehat{\mu}_i(t)}),
\label{equ:trans}
\end{aligned}
\end{equation}
subject to (\ref{equ:mu}), where $\widehat{\tilde{\bm{\mu}}}(t)\!=\!(\widehat{\tilde{\mu}}(t), i \in \mathcal{\tilde{N}})$ collects the optimal solutions under the approximated values.
The above problem is still convex and can, therefore, be solved in a similar way as the problem introduced in Section IV-B.
As a result, the proposed algorithm can be easily extended to the scenario with partial and outdated feedback by simply substituting the objective function of (\ref{equ:powerdata}) in Algorithm~1 to (\ref{equ:trans}).

Although such an approximation will inevitably lead to optimality loss of overall system utility, rigorous analysis of the following theorem shows that the performance penalty can be compensated and asymptotically diminished by~increasing the control parameter $V$.
Let $\widehat{U}^*(\bm{v}^t)$ denote the system utility obtained by using approximation values (\ref{equ:Qa}) and (\ref{equ:Za}), it can be readily established that
\begin{theorem}
\label{lemma:O1Vage}
The gap between $\widehat{U}^*(\bm{v}^t)$, and $U^\star$ satisfies
\begin{equation}
U^\star - \widehat{U}^*(\bm{v}^t) \leq (B_1+B_2)/V
\end{equation}
where $B_2\!=\!\sum\nolimits_{i \in \mathcal{N}}m c_i^{\text{max}}[(1+\frac{1}{p^2})c_i^{\text{max}}+2A_i^{\text{max}}]$.
\end{theorem}
\begin{IEEEproof}
Please refer to Appendix E.
\end{IEEEproof}
In other words, the asymptotic optimality of the proposed approach under partial and outdated NSI is proved.

\section{Simulations}
In this section, we justify the analytical claims and evaluate the performance of our proposed algorithm based on the simulation settings in~\cite{bibli:WMEC1,bibli:protocol2}, and~\cite{bibli:myIoT}.
The considered IIoT network includes ten WDs, whose distances from the AP is set to be  $d_i\!=\!3\!+\!(i\!-\!1)$ meters.
The wireless channel is modeled after the Rayleigh fading model and is set as $h_i(t)=10^{-3}d_i^{-\alpha}\widetilde{h}(t)$~\cite{bibli:protocol2}, where $\alpha$ denotes the path-loss exponent, $\widetilde{h}(t)$ is an exponentially distributed random variable with unit mean that represents the short-term fading.
As suggested in~\cite{bibli:myIoT}, we assume $\alpha=2$, $W=0.2$MHz, and $N_0=10^{-9}W$.
For any WD~$i$, $\xi_i\!=\!0.8$ and $A_i^{\text{max}}\!=\!1$Mb/s.
At the AP, the transmit power is set to be $P_0\!=\!2W$ and $r_i^{\text{max}}\!=\!50$kb/s.

The numerical results are obtained by averaging over 1000 independent realizations, in which $h_i(t)$, $A_i(t)$, and $r_i(t)$ are randomly selected according to the above assumptions at each time slot in every realization, modeling the real dynamics in IIoT.
It is worth noting that we neglect the computation time spent at the AP for conducting the proposed scheduling scheme, since the complexity of the schedule, as stated in Section IV-B, is low compared with CPU capability and does not explode with the network size $N$.
Even if the aforementioned time is non-negligible and accounts for constant proportion $\alpha$ of each time slot, a similar performance can be achieved by scaling the normalized time slot duration to $1\!+\!\alpha$.

For comparison purposes, we simulate two benchmark approaches: (a) proportional fair (PF)~\cite{bibli:channel}, where data offloading are scheduled based on proportional fairness; and~(b) homogeneous device optimization (HDO) \cite{bibli:myIoT}, where the system is optimized without considering data age.
Both of the benchmarks are measured in the absence of instantaneous network state information.
Besides, for self-contrast and simplicity, we refer to our proposed algorithm under complete and partial feedback as PCF and PPF, respectively.
The PCF obtains all devices' state information during its optimization, thus it provide the optimal bound of the PPF method.
The PPF with different data dropping price $p$ is also simulated to illustrate the adaptivity of the proposed algorithm.

\begin{figure}
\centering
\subfigure[]{
\label{fig:throughput} 
\includegraphics[width=0.49\linewidth]{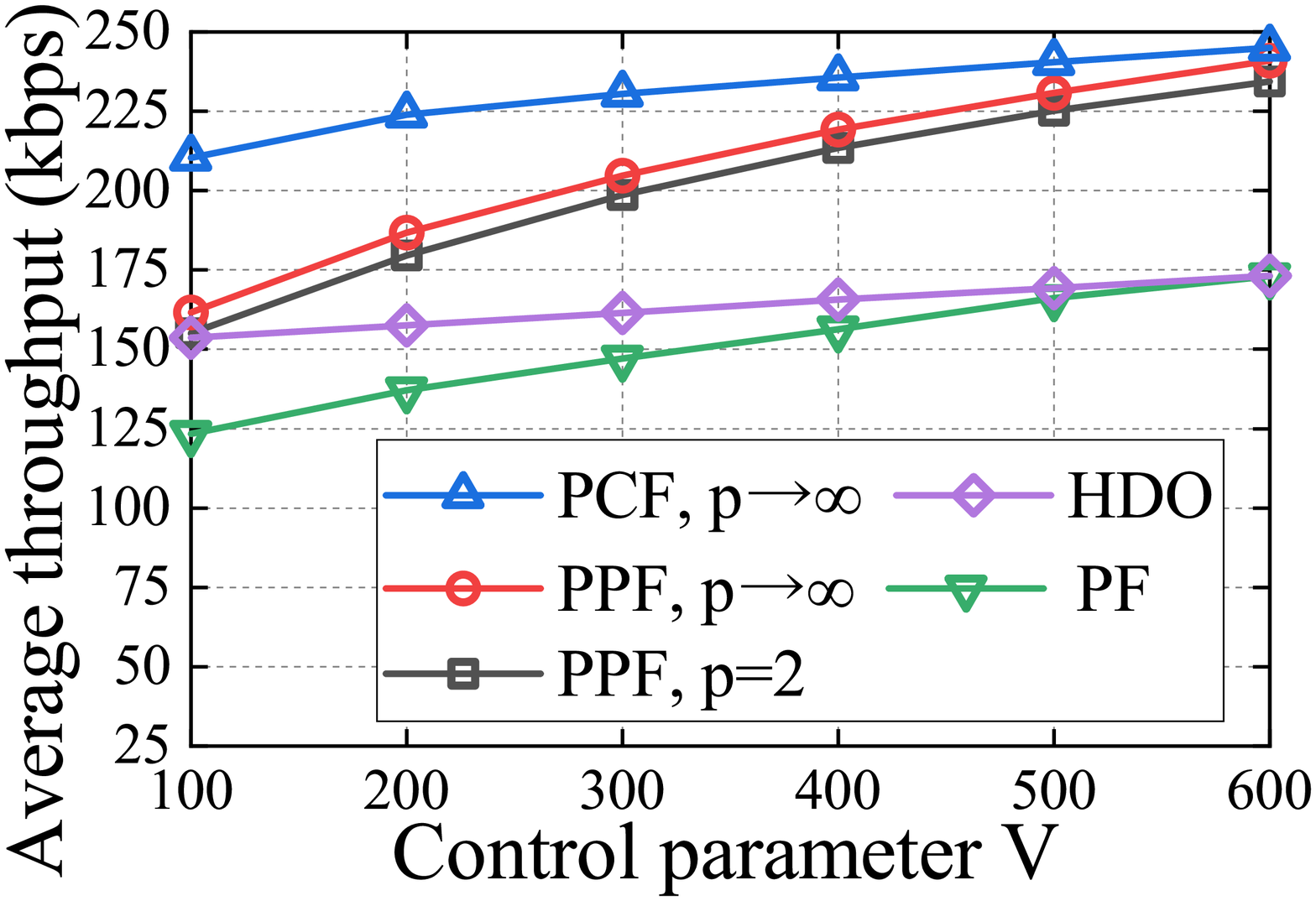}}
\hspace{-0.15in}
\subfigure[]{
\label{fig:fairness} 
\includegraphics[width=0.49\linewidth]{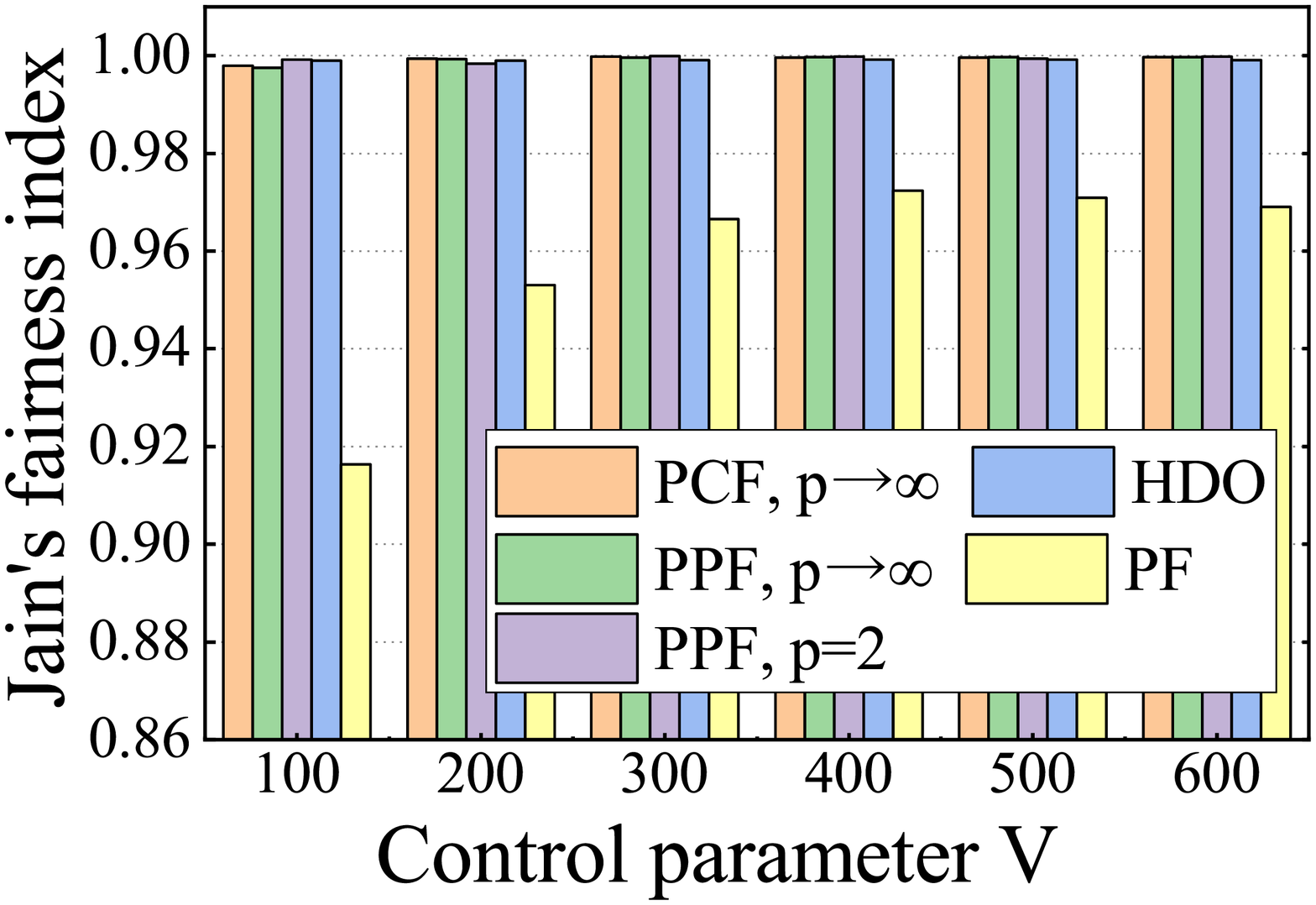}}
\caption{Impact of $V$ on (a) average system throughput and (b) Jain's fairness index of the proposed approach and benchmarks.}
\label{fig:simulation1} 
\end{figure}

Fig. \ref{fig:simulation1} illustrates the throughput and fairness (measured by Jain's index~\cite{bibli:channel}) against $V$.
The proposed algorithm under either complete feedback (i.e., PCF) or partial feedback (i.e. PPF) achieves higher throughput with a high fairness guarantee than PF and HDO.
The poor performance of PF and HDO with large $V$ suggests that $Z_i(t)$ help promote data offloading, which results from WDs getting more opportunities to offload, as can be seen from (\ref{equ:exclu}).
Besides, the decreasing slopes of the three curves of the proposed algorithm under different conditions in Fig.~\ref{fig:throughput} coincide with what revealed in Theorem~\ref{lemma:O1V}.
Moreover, the gaps among the three curves also verify the impacts of partial feedback and data discard price on system utility, as shown in Theorem~\ref{lemma:O1Vage}.

\begin{figure}
\centering
\subfigure[]{
\label{fig:age} 
\includegraphics[width=0.53\linewidth]{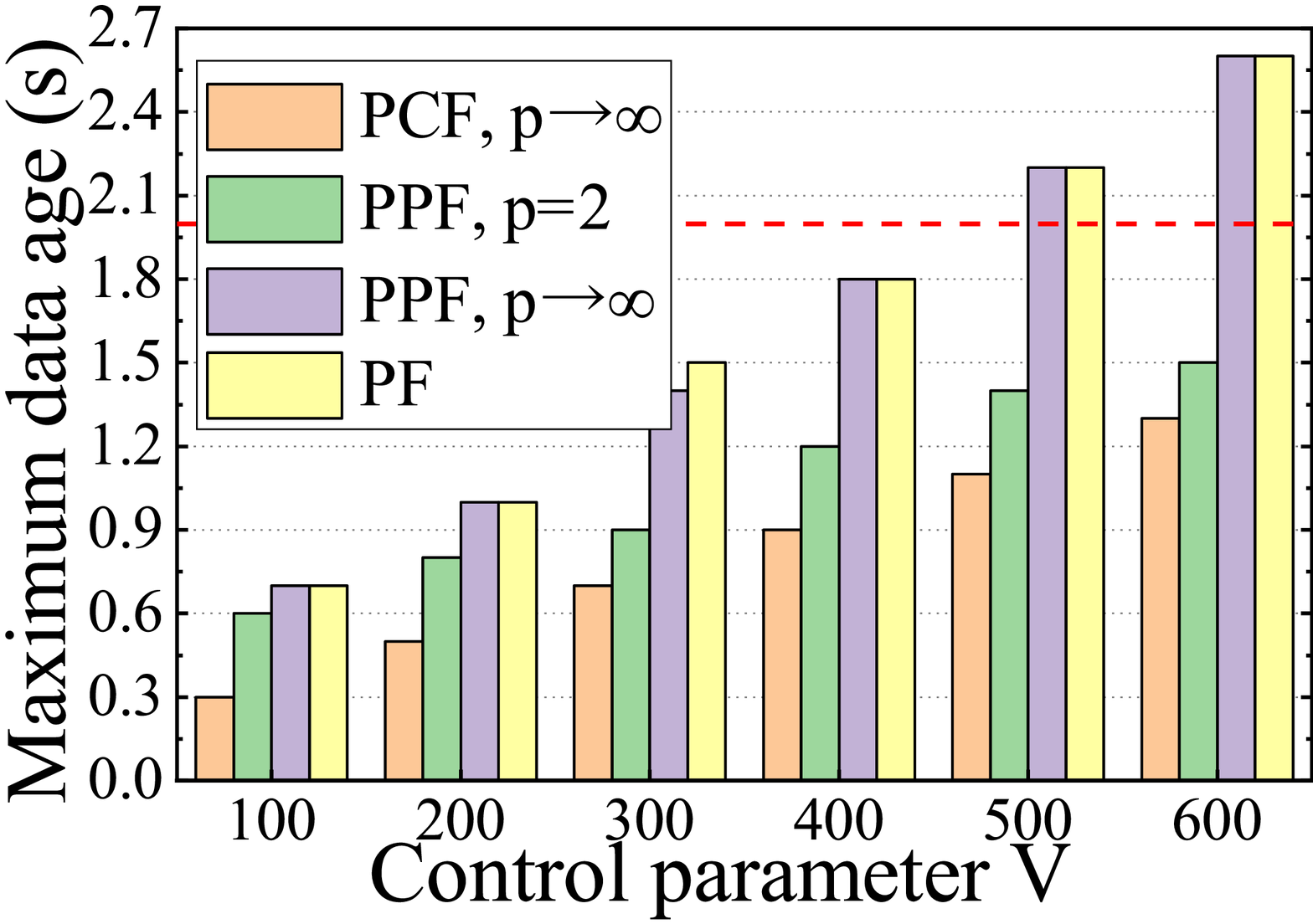}}
\subfigure[]{
\label{fig:age2} 
\includegraphics[width=0.33\linewidth]{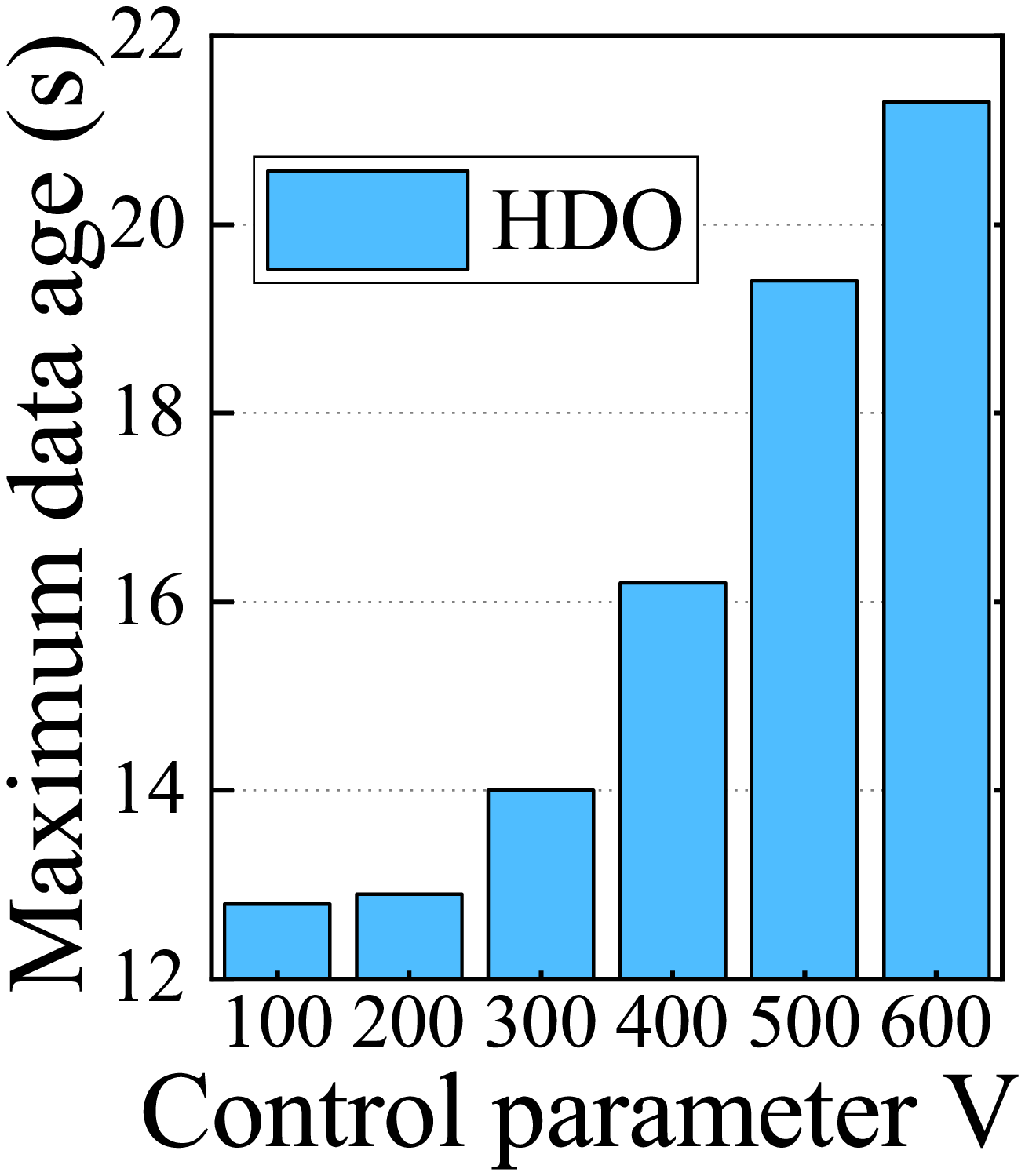}}
\caption{Maximum age of data from WDs versus $V$ using (a) proposed approach under different conditions and PF, and (b) HDO.}
\label{fig:simulation2} 
\end{figure}

Fig. \ref{fig:simulation2} shows the impact of $V$ on maximum data age.
Both PCF and PPF with $p=2$ outperform PF and HDO under different $V$, demonstrating the effectiveness of the proposed algorithm.
Even with $p\! \to \!\infty$, which means refusal to data dropping, PPF maintains its data as fresh as PF.
This is achieved by meticulously designing the virtual queues $\bm{Z}(t)$ to embedded data age into the optimization.
We also observe that the increases of $V$ and $p$ lead to the ``older" of offloaded data under our proposed algorithm, which agrees with Theorem \ref{lemma:OV}.
Meanwhile, the results provide guidelines for choosing applicable parameters to meet the demands in real implementations: one should select appropriate $V$ and $p$ to maximize the average throughput under the data age constraint.
For example, if the system requires data age to be less than $2s$, i.e., $g_{\text{max}}\!=\!2s$ as shown by the red line in Fig.~\ref{fig:age}, $V$ can be set as any value that is not larger than 400 for PPF with $p\! \to \!\infty$.
However, $V\!=\!400$ seems to be the best choice since it provides higher average throughput than that of smaller $V$, as can be seen from Fig. 2.

\begin{figure}
\centering
\subfigure[]{
\label{fig:Qlen} 
\includegraphics[width=0.49\linewidth]{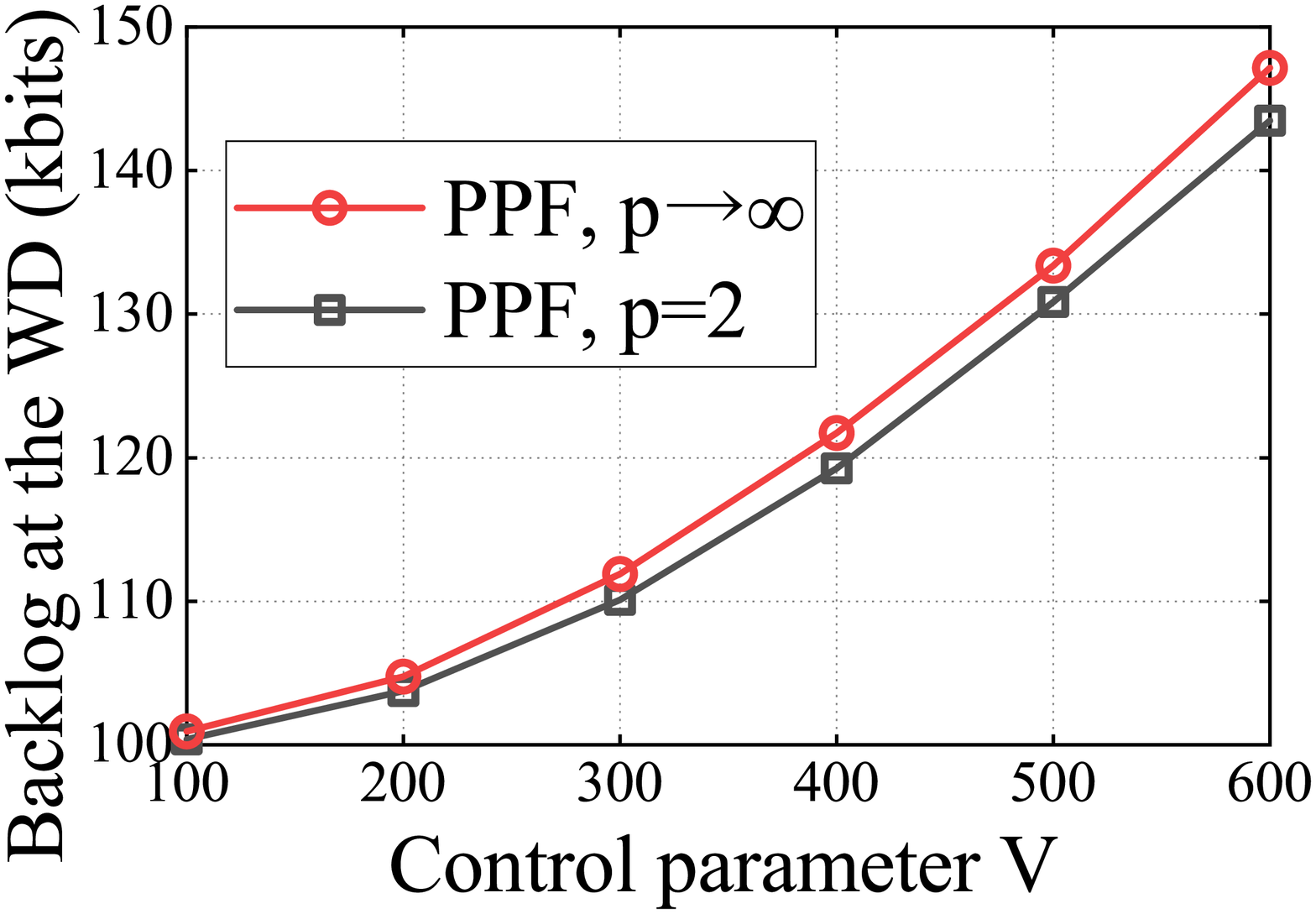}}
\hspace{-0.15in}
\subfigure[]{
\label{fig:Slen} 
\includegraphics[width=0.49\linewidth]{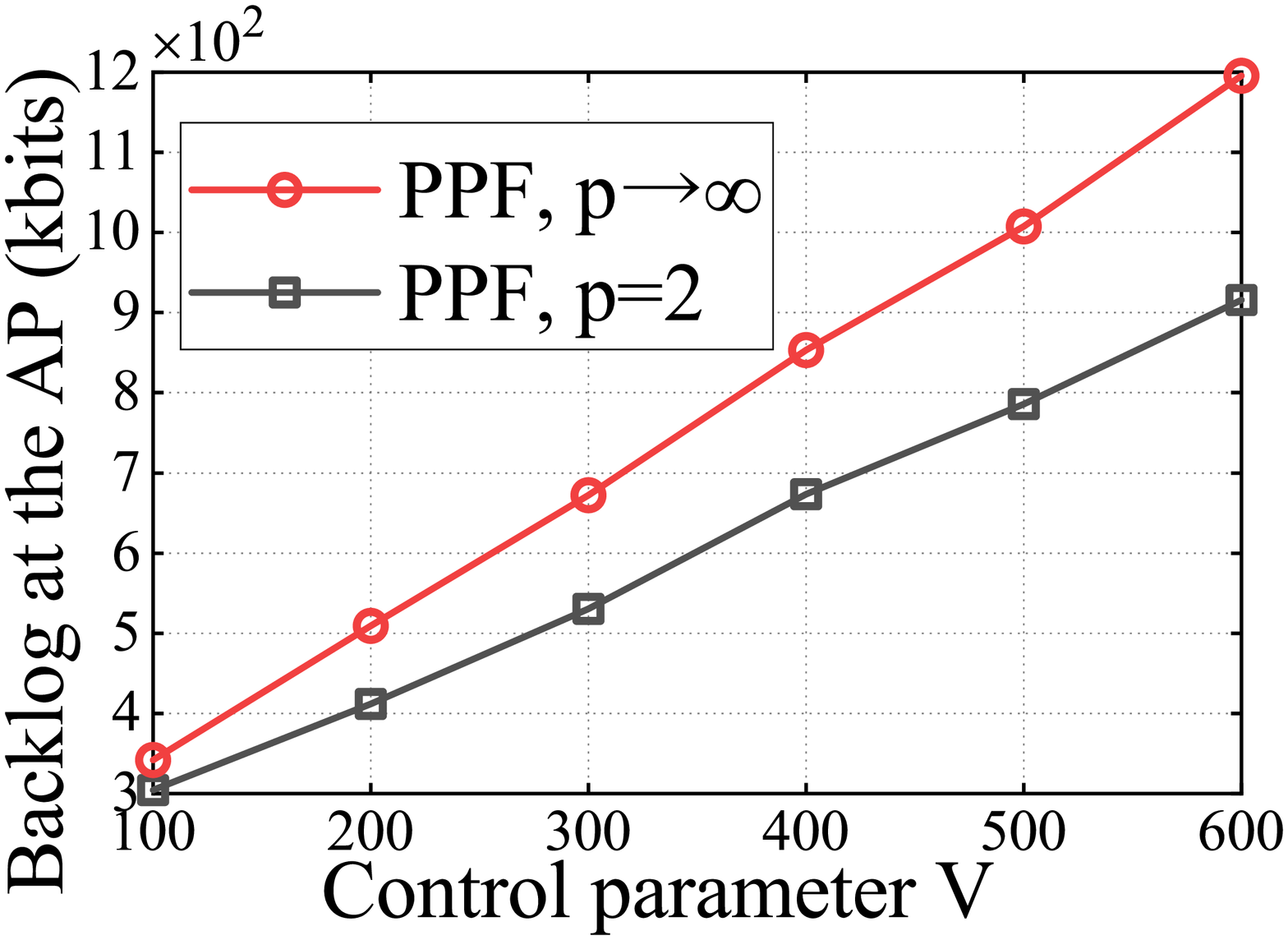}}
\caption{Impact of $V$ on maximum data backlog at the (a) WD and (b) AP.}
\label{fig:length} 
\end{figure}

Fig. \ref{fig:length} compares the data backlogs at WDs and the AP with different $V$ and $p$.
It can be observed that the maximum data backlogs at both sides grow as the control parameter $V$ increases, which corroborate their upper bounds given in Theorem \ref{lemma:OV}, thus guaranteeing system stabilities considered in problem~\textbf{P1}.
The reason why the curves in Fig. \ref{fig:Qlen} do not show a linear characteristics when $V$ increases from 100 to 300 lies in that WDs need a certain amount of data to satisfy the offloading condition $Q_i(t)\!+\!\frac{1}{p}Z_i(t)\!\geq \!S_i(t)$.
In Fig.~\ref{fig:length}, we also see that a higher data dropping price gives rise to backlogs at the devices, which is because a large $p$ encourages the maintaining of stale data.

The complexity of the proposed algorithm comes from two aspects.
For WDs, the optimal solutions of problems (14) and (15) can be obtained with closed-form by convex optimization technique with $\mathcal{O}(1)$ complexity, which is the same with PF and HDO.
The low complexity is thus very suitable for WDs with limited computing capacity in IIoT.
For the AP, the complexity to allocate time portions is at most with the order $\mathcal{O}(N)$ln($1/\kappa$)ln($1/\sigma$) as discussed in Section IV-B.
Compared to PF with $\mathcal{O}(N^2)$ complexity~\cite{bibli:channel} and HDO with a maximal computational complexity order of $\mathcal{O}(N^3)$~\cite{bibli:myIoT}, the proposed algorithm is more manageable in large-size IIoT networks where the network size $N$ dominates the overall complexity.

\section{Conclusions}
In this paper, we studied the dynamic resource schedules for wireless powered MEC in IIoT under partial outdated network knowledge.
The overall system utility consisting of throughput, device fairness, and data age was maximized by jointly optimizing wireless energy transfer, data collection, discard, and offloading.
Capitalizing on the Lyapunov optimization technique, an online algorithm was developed to make asymptotically optimal control decisions of the challenging problem.
The closed and semi-closed expressions for the optimal decisions were derived by convex optimization and Lambert W function.
Moreover, the optimality loss due to non-available real-time system information was proved to be bounded and asymptotically diminished.
Extensive simulations verified our theoretical analysis and the efficiency of the proposed approach.

The integration of the multi-antenna APs to the wireless powered MEC network is promising to increase the input power of the energy harvester, therefore enhancing the transmission efficiency.
In such a scenario, the proposed algorithm is still meaningful.
We can adopt the energy harvesting model that is commonly used in the multi-antenna WPT literature~\cite{bibli:delay2} with the assumption that the input radio frequency power is within the linear regime of the device's rectifier.
The energy harvested model can thus be transformed into a form that is proportional to the WPT time and can then be applied to the proposed approach.
By this means, energy efficiency can be boosted and the amount of offloaded data can also be increased.
However, this strategy requires an appropriate channel state information acquisition scheme for WPT.
How to optimally exploit the channel frequency diversity and the beamforming gain leaves for future work.

\appendices
\section{}
By exploiting Lyapunov optimization, we first define a quadratic Lyapunov function as $L(t)\!=\!\frac{1}{2}\sum\nolimits_{i \in \mathcal{N}}[Q_i^2(t)\!+\!S_i^2(t)\!+\!\frac{1}{p^2}Z_i^2(t)]$.
The drift-plus-penalty expression can thus be given by
\begin{equation}
\Delta_V(t)\!=\!\mathbb{E}\big\{L(t\!+\!1)\!-\!L(t)\!-\!VU(\bm{v}^t)\!\mid\!\bm{\Theta}(t)\big\}
\label{equ:drift}
\end{equation}
where $\bm{\Theta}(t)\! =\![\bm{Q}(t),\bm{S}(t),\bm{Z}(t)]$ is the concatenated vector of system queues.

Taking squares on both sides of (\ref{equ:Qt}), leveraging the identity inequality $([x\!-\!y]^+ \!+z)^2 \leq x^2\! + \!y^2\! +\! z^2\! +\! 2x(z\!-\!y)$ for any $x, y, z\!\geq\!0$, and summing over $i\in \mathcal{N}$, we obtain
\begin{equation}
\begin{aligned}
&\frac{1}{2}\sum\nolimits_{i \in \mathcal{N}}\big[Q_i^2(t\!+\!1)\!-\!Q_i^2(t)\big] \leq \frac{1}{2}\sum\nolimits_{i \in \mathcal{N}}\big[\big(c_i(t)\!+\!d_i(t)\big)^2\\
&\!+\!a_i^2(t)\big]\!+\!Q_i(t)\big[a_i(t)\!-\!c_i(t)\!-\!d_i(t)\big],
\label{equ:C4}
\end{aligned}
\end{equation}
Similarly, by using (\ref{equ:Ft}) and (\ref{equ:Zt}), we have
\vspace*{-5pt}
\begin{equation}
\begin{aligned}
&\frac{1}{2}\sum\nolimits_{i \in \mathcal{N}}[S_i^2(t\!+\!1)\!-\!S_i^2(t)] \leq \frac{1}{2}\sum\nolimits_{i \in \mathcal{N}}[r_i^2(t)\!+\!c_i(t)^2]\\
&+\sum\nolimits_{i \in \mathcal{N}}S_i(t)[c_i(t)\!-\!r_i(t)],
\label{equ:C5}
\end{aligned}
\vspace*{-5pt}
\end{equation}
and
\begin{equation}
\begin{aligned}
&\frac{1}{2p^2}\sum\nolimits_{i \in \mathcal{N}}\big[Z_i^2(t\!+\!1)\!-\!Z_i^2(t)\big]
\leq \frac{1}{2p^2}\sum\nolimits_{i \in \mathcal{N}}\big[p \epsilon_i\!-\!\frac{1}{p}c_i(t)\\
&\!-\!p d_i(t)\big]^2
+Z_i(t)[p\epsilon_i\!-\!\frac{1}{p}c_i(t)\!-\!p d_i(t)]
\label{equ:C3}
\end{aligned}
\end{equation}

Substituting (\ref{equ:C4})-(\ref{equ:C3}) into (\ref{equ:drift}), taking conditional expectation on $\bm{\Theta}(t)$, subtracting $V \mathbb{E}\big\{U(\bm{v}^t)\!\mid\! \bm{\Theta}(t) \! \big\}$, and using the boundedness assumptions (\ref{equ:a}), (\ref{equ:d}), $c_i(t)\!\leq\!c_i^{\text{max}}$, and $r_i(t)\!\leq\!r_i^{\text{max}}$ yield
\begin{equation}
\begin{aligned}
&\Delta_V(t) \leq B_1\!-\!V \mathbb{E}\big\{U(\bm{v}^t)\!\mid\! \bm{\Theta}(t) \! \big\}\!+\!\sum\nolimits_{i \in \mathcal{N}}Q_i(t)\mathbb{E}\big\{a_i(t)\\
&\!-\!c_i(t)\!-\!d_i(t)\!\mid\! \bm{\Theta}(t) \!\big\}\!+\!\sum\nolimits_{i \in \mathcal{N}}S_i(t)\mathbb{E}\big\{c_i(t)\!-\!r_i(t)\!\mid\! \bm{\Theta}(t) \!\big\}\\
&\!+\!\sum\nolimits_{i \in \mathcal{N}}Z_i(t)\big\{p\epsilon_i\!-\!\frac{1}{p}c_i(t)\!-\!d_i(t)\!\mid\! \bm{\Theta}(t) \!\big\}
\end{aligned}
\label{equ:unequal}
\end{equation}
where $B_1\!=\!\frac{1}{2}\max\{\epsilon_i^2,(A_i^{\text{max}}\!+\!\frac{1}{p^2}c_i^{\text{max}}\!-\!\epsilon_i)^2\}
\!+\!\frac{1}{2}[(c_i^{\text{max}}\!+\!A_i^{\text{max}})^2\!+\!(A_i^{\text{max}})^2\!+\!(c_i^{\text{max}})^2\!+\!(r_i^{\text{max}})^2]$.

According to the principle of opportunistically minimizing an expectation~\cite{bibli:Lyapunov}, the original problem can be transformed to minimize (\ref{equ:unequal}), subject to instantaneous constraints (\ref{equ:mu}), (\ref{equ:a}), and (\ref{equ:d}).
Rearranging (\ref{equ:unequal}), and suppressing $B_1$ and $r_i(t)$ that are independent of the optimization variables prove the results.

\section{}
Problem (19) is a convex optimization problem, which can be solved by Lagrangian method given by
\begin{equation}
\begin{aligned}
L(\tilde{\bm{\mu}}_c(t),\lambda(t))=&-\sum\limits_{i \in \mathcal{N}\backslash \mathcal{N}_t}D_i(t) \mu_i(t) \log_{2}(1+\frac{\delta_i(t)\mu_0(t)}{\mu_i(t)})\\ &+\lambda(t)\left[\sum\nolimits_{i\in \mathcal{\tilde{N}}\backslash \mathcal{N}_t} \mu_i(t)-1\right]
\end{aligned}
\end{equation}
where $\lambda(t)\!\geq\!0$ denotes the Lagrange multiplier associated with the constrains in (\ref{equ:exclu}), and $D_i(t)\!=\!-[S_i(t)\!-\!Q_i(t)\!-\!\frac{1}{p}Z_i(t)]W$.
The constraints in (\ref{equ:exclu}) are all linear inequalities, thus refined Slater condition holds.
Moreover, strong duality holds since problem (\ref{equ:exclu}) is convex.
Therefore, the Karush-Kuhn-Tucker (KKT) conditions provide necessary and sufficient conditions for the global optimality of problem (\ref{equ:exclu}) \cite{bibli:convex}, which satisfy
\begin{equation}
\lambda^{*}(t)\left[\sum\nolimits_{i\in \mathcal{\tilde{N}}\backslash \mathcal{N}_t} \mu_i^{*}(t)-1\right]=0,
\label{equ:sum-one}
\end{equation}
\begin{equation}
\frac{\partial L}{\partial \mu_0(t)}\!=\!-\!\sum\limits_{i \in \mathcal{N}\backslash \mathcal{N}_t} \frac{D_i(t)}{\ln2}\frac{\delta_i(t)}{1\!+\!\frac{\delta_i(t)\mu_0^{*}(t)}{\mu_i^{*}(t)}}\!+\!\lambda^{*}(t)\!=\!0,
\label{equ:KKT2}
\end{equation}
\begin{equation}
\frac{\partial L}{\partial \mu_i(t)}= -\frac{D_i(t)}{\ln2}\Xi(\frac{\delta_i(t)\mu_0^{*}(t)}
{\mu_i^{*}(t)})+\lambda^{*}(t)=0,~i \in \mathcal{N}\backslash \mathcal{N}_t,
\label{equ:KKT3}
\end{equation}
where $\lambda^{*}(t)$ is the optimal Lagrangian multiplier in time slot $t$, $\Xi(x)$ is a monotonically increasing function defined as
\begin{equation}
\Xi(x) \triangleq \ln (1+x) + \frac{1}{1+x} -1.
\end{equation}

Note from (\ref{equ:sum-one}) that $\sum\nolimits_{i\in \mathcal{\tilde{N}}_c} \mu_i^{*}(t)\!=\!1$ must hold, otherwise we can allocate the remain time to $\mu_0(t)$ to further improve the transmission power of all devices. Meanwhile, we can also infer that $\lambda^{*}(t)\!>\!0$ holds strictly.

For arbitrary $i\in \mathcal{N}\backslash \mathcal{N}_t$, from (\ref{equ:KKT3}) we have
\begin{equation}
\ln \left(1+\frac{\delta_i(t)\mu_0^{*}(t)}
{\mu_i^{*}(t)}\right) + \frac{1}{1+\frac{\delta_i(t)\mu_0^{*}(t)}
{\mu_i^{*}(t)}}
=\frac{\ln2\lambda^{*}(t)}{D_i(t)}+1
\end{equation}
By dividing -1 and then taking a natural exponential operation at both sides, we have
\begin{equation}
\frac{-1}{1\!+\!\frac{\delta_i(t)\mu_0^{*}(t)}
{\mu_i^{*}(t)}}\text{exp}\left(\frac{-1}{1\!+\!\frac{\delta_i(t)\mu_0^{*}(t)}
{\mu_i^{*}(t)}}\right) \!=\!-\! \text{exp}\left(\frac{-\ln2\lambda^{*}(t)}{D_i(t)}\!-\!1\right)
\end{equation}
According to the Lambert function~\cite{bibli:WMEC1}, we obtain
\begin{equation}
\frac{\delta_i(t)\mu_0^{*}(t)}
{\mu_i^{*}(t)}=-\left[ 1+\frac{1}{W_0(-\text{exp}({-\frac{\ln2\lambda^{*}(t)}{D_i(t)}-1}))}  \right]
\label{equ:ratio}
\end{equation}
Combining (\ref{equ:ratio}) and $\sum\nolimits_{i\in \mathcal{\tilde{N}}_c} \mu_i^{*}(t)\!=\!1$ leads to the results in Lemma \ref{lemma:mu}.

\section{}
According to [21, Th. 4.5], there exists a stationary optimal policy $\Pi^\star$ that achieves optimal utility $U^\star$ while maintaining system stability.
Therefore, for $\Pi^\star$, each term on the RHS of (\ref{equ:unequal}) is non-positive due to queue constraints, and we have
\begin{equation}
\Delta_V(t)\!\leq\! B_1\!-\!VU^\star
\label{equ:E3}
\end{equation}
Taking iterated expectations of (\ref{equ:E3}) and telescoping sums over $t$, dividing both sides by $Vt$, and rearranging terms yields
\begin{equation}
\frac{1}{t}\sum\nolimits_{\tau=0}^{t-1}\mathbb{E}\big\{U(\bm{v}^t)\!\big\} \!\geq\! U^\star\!-\frac{B_1}{V}\!-\frac{\mathbb{E}\{L(0)\}}{Vt}
\label{equ:E4}
\end{equation}
Given $\mathbb{E}\{L(0)\}\! \leq \!\infty$, taking limits in (\ref{equ:E4}) as $t\! \to\! \infty$ concludes the proof.

\section{}
We first prove (\ref{equ:Qmax}) through mathematical induction.
It clearly holds for $t\!=\!0$ as $Q_i(0)\!=\!0$.
Suppose the upper bound holds at slot $t$.
Then, if $Q_i(t)\!\leq \!V(2\!-\!e^{-p})$, the difference between $Q_i(t)$ and $Q_i(t\!+\!1)$ is less than $A_i^{\text{max}}$, hence, we have $Q_i(t\!+\!1)\!\leq\! Q_i(t)\!+\!A_i^{\text{max}}$.
Otherwise, if $Q_i(t) > V(2\!-\!e^{-p})$, $a_i(t)\!=\!0$ according to (\ref{equ:condition1}).
Thus, $Q_i(t)$ cannot increase at slot $t$, i.e., $Q_i(t\!+\!1)\!\leq\!Q_i(t)$.
As a result, the upper bound also holds at slot $t\!+\!1$, which completes the proof of~(\ref{equ:Qmax}).

Likewise, we can prove (\ref{equ:Zmax}) through mathematical induction.
Since (\ref{equ:Zmax}) holds at slot $t\!=\!0$, we suppose it also holds at slot $t$.
According to (\ref{equ:drops}), if $Z_i(t) > pV$, we get $d_i(t)\!=\!A_i^{\text{max}}$.
As $\epsilon_i\leq A_i^{\text{max}}$, based on (\ref{equ:Zt}), we have $Z_i(t\!+\!1)\!\leq \!Z_i(t)\! \leq\! p(V\!+\!\epsilon_i)$.
If $Z_i(t)\!\leq \!p V$, we have $Z_i(t\!+\!1)\! \leq\! p(V\!+\!\epsilon_i)$ by considering the maximum arrival rate.
In other words, (\ref{equ:Zmax}) holds at slot $t\!+\!1$.
This concludes the proof of (\ref{equ:Zmax}).

Next, we can prove (\ref{equ:Smax}) by using the conclusions above.
Again, it holds for $t\!=\!0$ as $S_i(0)\!=\!0$.
We assume it also holds at slot $t$.
Consider the case when $S_i(t)\leq Q_i^{\text{max}}+\frac{1}{p}Z_i^{\text{max}}$, we have $S_i(t\!+\!1)\!\leq\! S_i(t)\!+\!c_i^{\text{max}}$ by substituting the maximum link capacity.
Otherwise, if $S_i(t)\! > \!Q_i^{\text{max}}+\frac{1}{p}Z_i^{\text{max}}$, no data is allowed to transmitted according to the offloading condition in Section IV-B, i.e., $c_i(t)\!=\!0$.
Hence, $S_i(t\!+\!1) \!\leq\!S_i(t)$.
To sum up, (\ref{equ:Smax}) holds at $t\!+\!1$, thus concludes its upper bound proof.

Finally, we proceed to verify (\ref{equ:gmax}) by proof of contradiction.
Suppose (\ref{equ:gmax}) is not true, i.e., there exists data whose age exceeds $g_i^{\text{max}}$ for $i \!\in \!\mathcal{N}$.
Since collected data is either offloaded or dropped in a FIFO manner, the sum of $c_i(\tau)+d_i(\tau)$ over $\tau \in \{t\!+\!1,\cdot \cdot \cdot, t\!+\!g_i^{\text{max}}\}$ should be less or equal to $Q_i(t\!+\!1)$.
Otherwise, all data admitted at slot $t$ would be cleared before the $g_i^{\text{max}}$ expires.
Therefore, we obtain
\begin{equation}
\sum\nolimits_{\tau=t\!+\!1}^{t\!+\!g_i^{\text{max}}} [c_i(\tau)\!+\!d_i(\tau)] < Q_i(t\!+\!1)\! \leq \!Q_i^{\text{max}}
\label{equ:B1}
\end{equation}
For any $\tau \in \{t\!+\!1,\cdot \cdot \cdot, t\!+\!g_i^{\text{max}}\}$, from (\ref{equ:Zt}), we have
\begin{equation}
\begin{aligned}
Z_i(\tau+1)& \geq Z_i(\tau)-\frac{1}{p} c_i(\tau)-p d_i(\tau)+p \epsilon_i\\
& \geq Z_i(\tau)-p [c_i(\tau)+ d_i(\tau)]+p \epsilon_i
\end{aligned}
\label{equ:Zunequl}
\vspace*{-5pt}
\end{equation}
Summing (\ref{equ:Zunequl}) over $\tau$ by telescoping sums, using the fact that $0 \!\leq \!Z_i(\tau)\! \leq \!Z_i^{\text{max}}$, and rearranging the terms yields
\begin{equation}
g_i^{\text{max}}\epsilon_i \leq \sum\nolimits_{\tau=t\!+\!1}^{t\!+\!g_i^{\text{max}}} [c_i(\tau)\!+\!d_i(\tau)] +  \frac{1}{p}Z_i^{\text{max}}
\label{equ:B2}
\end{equation}
Combining (\ref{equ:B1}) and (\ref{equ:B2}), and then dividing both sides by $\epsilon_i$, we have
\begin{equation}
g_i^{\text{max}} < \frac{1}{\epsilon_i}(Q_i^{\text{max}} +  \frac{1}{p}Z_i^{\text{max}})
\label{equ:B3}
\end{equation}
which contradicts (\ref{equ:gmax}).
This concludes the proof.

\section{}
According to (\ref{equ:Qt}) and (\ref{equ:Qa}), the difference between the approximate and the actual data backlogs satisfies $\widehat{Q}_i(t)\!-\!Q_i(t)\leq[c_i(t)\!+\!d_i(t)]\tau_i^Q\leq[c_i^{\text{max}}\!+\!A_i^{\text{max}}]m$.
Likewise, we can obtain  $\widehat{Z}_i(t)\!-\!Z_i(t)\leq[\frac{1}{p}c_i^{\text{max}}\!+\!pA_i^{\text{max}}]m$ from (\ref{equ:Zt}) and (\ref{equ:Za}).
Recall that $c_i(t)$ and $\widehat{c}_i(t)$ are the optimal channel allocation under the actual and approximate approaches, respectively.
Since $c_i(t)$ minimizes (13c), we can obtain that
\vspace*{-5pt}
\begin{equation}
\begin{aligned}
f_3(\tilde{\bm{\mu}}(t))=&\sum\nolimits_{i \in \mathcal{N}}[S_i(t)-Q_i(t)-\frac{1}{p}Z_i(t)]c_i(t) \\
\leq &\sum\nolimits_{i \in \mathcal{N}}[S_i(t)-Q_i(t)-\frac{1}{p}Z_i(t)]\widehat{c}_i(t)
\end{aligned}
\vspace*{-3pt}
\end{equation}
Hence, we have
\begin{equation}
\begin{aligned}
&f_3(\tilde{\bm{\mu}}(t))-\widehat{f_3}(\widehat{\tilde{\bm{\mu}}}(t))\\
\leq&\sum\nolimits_{i \in \mathcal{N}}[\widehat{Q}_i(t)\!-\!Q_i(t)]\widehat{c}_i(t)\!+\!\frac{1}{p}\sum\nolimits_{i \in \mathcal{N}}[\widehat{Z}_i(t)\!-\!Z_i(t)]\widehat{c}_i(t)\\
\leq & \sum\nolimits_{i \in \mathcal{N}}m c_i^{\text{max}}[(1+\frac{1}{p^2})c_i^{\text{max}}+2A_i^{\text{max}}]=B_2
\end{aligned}
\label{equ:diff}
\vspace*{-3pt}
\end{equation}

The optimality loss brought by the approximation (\ref{equ:Qa}) and (\ref{equ:Za}) can thus be derived by
\begin{equation}
U^*(\bm{v}^t)-\widehat{U}^*(\bm{v}^t)=\frac{1}{V}\big[f_3(\tilde{\bm{\mu}}(t))-\widehat{f_3}(\widehat{\tilde{\bm{\mu}}}(t)) \big] \leq B_2/V.
\label{equ:gap2}
\end{equation}
Adding up (\ref{equ:gap1}) and (\ref{equ:gap2}) concludes the proof.

\ifCLASSOPTIONcaptionsoff
  \newpage
\fi



%

\bibliographystyle{IEEEtran}
\bibliography{Ref-MEC-TWC}

%




\end{document}